\documentclass[journal]{IEEEtran}
\usepackage{graphicx}
\usepackage{amsmath,amssymb}

\begin{document}
\title{Fine-pitch semiconductor detector for the \textit{FOXSI} mission}
%
% author names and IEEE memberships
% note positions of commas and nonbreaking spaces ( ~ ) LaTeX will not break
% a structure at a ~ so this keeps an author's name from being broken across
% two lines.
% use \thanks{} to gain access to the first footnote area
% a separate \thanks must be used for each paragraph as LaTeX2e's \thanks
% was not built to handle multiple paragraphs
%

\author{S. Ishikawa, S. Saito, H. Tajima, T. Tanaka, S. Watanabe, H. Odaka, T. Fukuyama, M. Kokubun, T. Takahashi, 
Y. Terada, S. Krucker, S. Christe, S. McBride and L. Glesener% <-this % stops a space
\thanks{Manuscript received -------- --, 2010. }% <-this % stops a space
\thanks{S. Ishikawa, S. Saito, S. Watanabe, H. Odaka, T. Fukuyama, S. Sugimoto and T. Takahashi are with Institute of Space and Astronautical Science, Japan Aerospace Exploration Agency, Sagamihara, Kanagawa, 
Japan, and also with Department of Physics, University of Tokyo, Bunkyo, Tokyo, Japan
 (e-mail: ishikawa@astro.isas.jaxa.jp).}%
 \thanks{M. Kokubun is with Institute of Space and Astronautical Science, 
 Japan Aerospace Exploration Agency, Sagamihara, Kanagawa, Japan.}%
\thanks{Y. Terada is with Department of Physics, Saitama University, Japan.}%
\thanks{H. Tajima and T. Tanaka are with Kavli Institute for Particle Astrophysics and Cosmology, Stanford University, Stanford, CA, USA.}%
\thanks{S. Krucker, S. McBride and L. Glesener are with the Space Science Laboratory, U.C. Berkeley, CA, USA.}%
\thanks{S. Christe is with NASA/Goddard Space Flight Center, MD, USA}%
}

\maketitle
\pagestyle{empty}
\thispagestyle{empty}

\begin{abstract}
The Focusing Optics X-ray Solar Imager (\textit{FOXSI}) is a NASA sounding rocket mission which will
study particle acceleration and coronal heating on the Sun through
high sensitivity observations in the hard X-ray energy band (5--15~keV).
Combining high-resolution focusing X-ray optics and fine-pitch imaging sensors,
\textit{FOXSI} will achieve superior sensitivity; two orders of magnitude better than
that of the \textit{RHESSI} satellite.  
As the focal plane detector, a Double-sided Si Strip Detector (DSSD) with a front-end ASIC (Application Specific Integrated Circuit) will 
fulfill the scientific requirements of spatial and energy resolution, low energy threshold and time resolution.  
We have designed and fabricated a DSSD with a thickness of 500~$\mu$m and a dimension of 9.6~mm$\times$9.6~mm, 
containing 128 strips with a pitch of 75~$\mu$m, which corresponds to 8~arcsec at the focal length of 2~m.  
We also developed a low-noise ASIC specified to \textit{FOXSI}.  
The detector was successfully operated in the laboratory at a temperature of $-$20$^\circ$C and with an applied bias voltage of 300~V, 
and the energy resolution of 430~eV at a 14~keV line was achieved.  
We also demonstrated fine-pitch imaging successfully by obtaining a shadow image, hence 
the implementation of scientific requirements was confirmed.  
\end{abstract}

%\begin{IEEEkeywords}
%IEEEtran, journal, \LaTeX, paper, template.
%\end{IEEEkeywords}

\section{Introduction}

\IEEEPARstart{I}{n}
solar flares, it is well known that electrons and ions are accelerated to high energies\cite{benz2008}.  
Accelerated particles emit hard X-rays (HXRs) by the bremsstrahlung process as they travel and lose their energy in the solar atmosphere.  Therefore, HXR observations of the Sun provide important information about the energy release process in solar flares.  Past solar HXR observations, such as those by the Hard X-ray Telescope onboard the 
\textit{Yohkoh} satellite
\cite{kosugi1991}
\cite{kosugi1992} and those by the Reuven Rematy High Spectroscopic 
Imager (\textit{RHESSI}) satellite\cite{lin2002}, have made use of non-focusing imaging techniques such as rotation modulation collimator imaging\cite{hurford2002}.  
However, rotation modulation collimators require image reconstructions which limit dynamic range, and also require
a large detector volume to obtain sufficient effective area, resulting in high background. 

\begin{figure}[!t]
\centering
\includegraphics[width=3.6in]{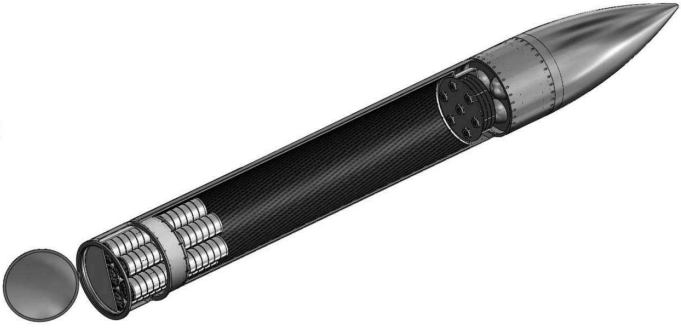}
\caption{Schematic of the \it{FOXSI} payload which consists of 7 telescope modules located at the optics plane (left) 
and 7 detectors at the the focal plane (right).  Various electronics are located behind the focal plane.  The focal length is 2~m.}
\label{fig_payload}
\end{figure}

Grazing-incidence HXR focusing optics are a promising new technology to avoid shortcomings of rotation modulation collimators. 
Focusing optics do not require image reconstructions and enable the use of small focal plane detectors while retaining large effective area, which imply a drastic reduction in non-solar background such as the one
due to cosmic-ray, thereby increasing the sensitivity to solar HXR sources.
A new sounding rocket mission, the Focusing Optics X-ray Solar Imager (\textit{FOXSI}) will test out grazing-incidence HXR focusing optics\cite{ramsey2002} combined with position-sensitive focal plane detectors for solar observations (Fig.~\ref{fig_payload}).  
\textit{FOXSI} will achieve superior sensitivity, two orders of magnitude better than that of \textit{RHESSI} around 10 keV.
The \textit{FOXSI} rocket is to be launched in October 2011.  

To meet the science objectives of the \textit{FOXSI} mission, a focal plane detector needs to satisfy the following requirements: 
A spatial resolution of $<$116~$\mu$m 
(corresponding to the angular resolution of the optics of $\sim$12~arcseconds), 
an energy resolution of below $\sim$1~keV (FWHM) and good photoabsorption efficiency (better than 50\%) in the 5--15~keV band (see Table~\ref{table_overview}).  
The focal plane detector also needs to be able to perform quasi-single photon counting for count rates up to $\sim$100~counts/s expected from the quiet Sun.  
Existing imaging technologies widely used in focusing X-ray optics do not meet all of the above requirements.
For example, charge-coupled devices can achieve good energy resolutions and low threshold energies, however, their time resolution is limited to an order of 1~s\cite{singh2005}.
Therefore, it is critical to develop a new focal plane detector for \textit{FOXSI}.

A Double-sided Silicon Strip Detector (DSSD) in combination with a low noise ASIC (Application Specific Integrated Circuit) is 
one of a few sensor technologies that could satisfy the requirements for the focal plane detector in \textit{FOXSI}.
We have developed a fine-pitch DSSD with a new detector structure to achieve 
the low noise requirement of this mission.  
In this paper, we present a detailed concept of the proposed solution, including experimental results to confirm its performance.

\begin{table}
\begin{tabular}{ll} \hline
Energy range & $\sim$5  to 15~keV\\
Energy resolution (FWHM) & $\sim$1~keV\\
Focal Length & 2~m\\
Angular resolution (FWHM) & 12~arcseconds \\
Field of view (HPD) & 640 $\times$ 640 arcseconds\\
Effective area & 120~cm$^2$ (8~keV), $\sim$10~cm$^2$ (15~keV)\\
Sensitivity & 0.004 cm$^{-2}$s$^{-1}$keV$^{-1}$($\sim$8~keV)\\
Dynamic range & 100 for source separation $>$30~arcseconds\\
Observation time & $\sim$360~s \\
Launch site & White Sands Missile Range, NM, USA \\
Launch date & late 2011 \\ \hline 
\end{tabular}
\caption{\textit{FOXSI} overview\cite{krucker2009foxsi}.}
\label{table_overview}
\end{table}

\section{Design of the \textit{FOXSI} Focal Plane Detector}

\subsection{Double-sided Silicon Strip Detector}
In order to take advantage of the good angular resolution of 12~arcseconds of the \textit{FOXSI} HXR focusing optics, we designed and fabricated a fine-pitch DSSD 
specific for \textit{FOXSI} (the \textit{FOXSI} DSSD).  
As the material for the semiconductor imaging detector, silicon has a sufficient photoabsorption efficiency in the energy band of interest.  
Although other materials such as cadmium telluride or cadmium zinc telluride have higher efficiencies and can achieve good performances at temperatures around $-$20$^\circ$C, 
silicon is a well-researched material available in high purity, and detectors with high performances can be obtained with high yield at low cost.  In addition, silicon detectors are quite radiation tolerant, 
and we have many years of 
experience in operating them in high radiation environments.

The \textit{FOXSI} DSSD has highly acceptor doped (p$^+$) silicon strips (p-side) and donor doped (n$^+$) silicon strips (n-side) implanted orthogonally on a n-type silicon wafer (n-bulk).  
Each n-side strip is surrounded by a floating p$^+$-doped implantation (p-stop) to insulate it from adjacent strips.  
Aluminum electrodes are directly coupled to each strip with an ohmic contact.  
DSSDs have been widely developed for astrophysical and nuclear physics applications\cite{sellin1992}\cite{tajima2003}.  
A DSSD with a strip pitch of 250~$\mu$m or 400~$\mu$m has been developed by our group as a scattering detector of a Si/CdTe Compton camera\cite{tajima2003, takeda2007, watanabe2007compton}. 
In this development, we added aluminum electrodes DC-coupled to p-stops in order to minimize the p-stop resistance since it generates Johnson noise.
A similar DSSD is also being developed for the Hard X-ray Imager on board the Japanese X-ray astronomy satellite \textit{ASTRO-H}\cite{takahashi2010, kokubun2010}.

\begin{figure}[!t]
\centering
\includegraphics[width=2.0in]{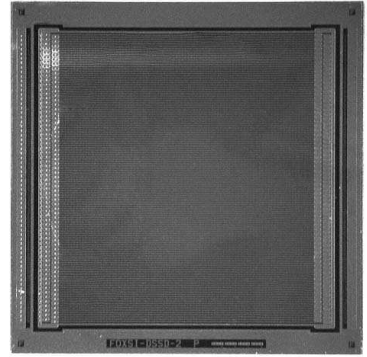}
\caption{Photograph of the 128$\times$128 \textit{FOXSI} DSSD (p-side is shown).  The active area is 9.6~mm $\times$ 9.6~mm, and the strip pitch is 75~$\mu$m.}
\label{fig_photo}
\end{figure}

The active area of the \textit{FOXSI} DSSD is 9.6~mm$\times$9.6~mm and the number of strips is 128 for p-side and n-side, providing position 
information for 128$\times$128$=$16384 pixels by reading out only 128$+$128 = 256~channels, resulting in lower power consumption.  
The pitch of the strips is 75~$\mu$m, corresponding to an angular resolution of 8~arcseconds at the focal length of 2~m.  
Hence, the spatial resolution of the optics is oversampled by a factor of 1.5.
The thickness is 500~$\mu$m, which implies a photoabsorption efficiency of 98~\% for 10~keV and 68~\% for 15~keV.  
Guard-ring electrodes with a width of 100~$\mu$m are implemented on both sides to block the leakage current from the periphery.  
Fig.~\ref{fig_photo} shows a photo, and Table~\ref{table_dssd} shows the specifications of the \textit{FOXSI} DSSD.  
The device is manufactured by Hamamatsu Photonics, Japan.  
\begin{table}
\begin{tabular}{ll} \hline
Active Area & 9.6~mm$\times$9.6~mm\\
Thickness & 500~$\mu$m\\
Strip pitch & 75~$\mu$m\\
Number of strips per side & 128\\ \hline
\end{tabular}
\caption{Specifications of the \textit{FOXSI} DSSD.}
\label{table_dssd}
\end{table}

\subsection{The Low Noise Front End ASIC VATA451}
\begin{figure*}[!t]
\centering
\includegraphics[width=7in]{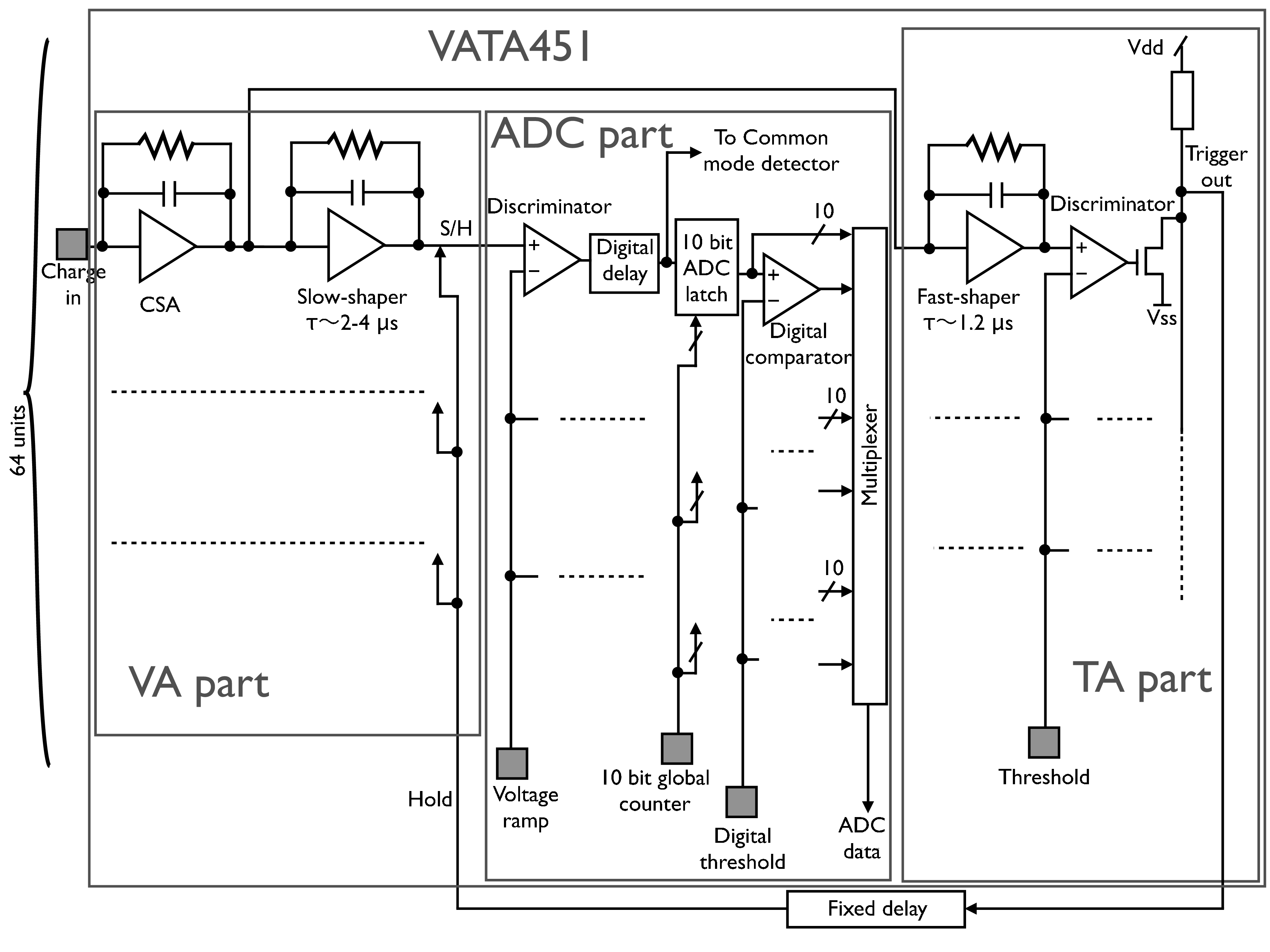}
\caption{Functional block diagram of the readout ASIC, VATA451.  
The ASIC consists of a pulse-shaping VA part and a trigger-generating TA parts.  
An ADC is also included.}
\label{fig_vatablock}
\end{figure*}

To achieve good energy resolutions and a low threshold energy, a 64-channel analog ASIC, VATA451, has been developed by ISAS, KIPAC and GM-IDEAS based on a low noise front end 
ASIC previously developed for Compton cameras for applications including \textit{ASTRO-H}\cite{tajima2004vata}.
Fig.~\ref{fig_vatablock} shows a function block diagram for VATA451.
Each channel of the analog circuit of the VATA451 consists of a charge-sensitive amplifier (CSA) 
followed by two shaping amplifiers, one with a fast shaping time for generating triggers (TA section), and the other with a slow shaping time for charge measurements (VA section).
If the output signal from the fast shaper of one or more channels is greater than a threshold value in the TA section, the output signals from the slow shaper of each channel will be sampled and held with an appropriate delay.
A Wilkinson-type analog-to-digital converter, by which all channels can be digitized in parallel, is also included in the VATA451.  
The ADC is 10 bit and takes 100~$\mu$s for digitization with a 10 MHz clock, and output signals are multiplexed.
A common-mode noise calculator is also built into the ASIC.  
Common-mode noise, which is a noise component common to all channels of an ASIC for a given event, can be obtained by using the median value of the signals from all channels for that event.  
These analog and ADC architectures are functionally the same as the former ASIC, VA32TA6\cite{watanabe2009}.

This ASIC is specifically optimized for 
the \textit{FOXSI} mission as follows.
The gain of the CSA is higher than that of \textit{ASTRO-H} to achieve better noise performance since the energy range of \textit{FOXSI} (below 15~keV) is lower than that of \textit{ASTRO-H}.
The fast shaper provides a longer shaping time constant option for generating triggers, which will be beneficial to achieve a low threshold energy required by \textit{FOXSI}.
The time constant of the slow shaper can be adjusted from 2~$\mu$s to 4$\mu$s, and that of the fast shaper can be set to 0.6~$\mu$s or 1.2~$\mu$s.  

The input FET of the VATA451 is optimized to minimize the noise for an input capacitance of 5~pF and a leakage current of 10~pA within a power constraint of 1~mW/channel, resulting in an equivalent noise charge (ENC) of 64~$e^-$ (RMS) for such input loads with a shaping time of 3~$\mu$s.

Since there are 128 strips for each p-side and n-side, 
two ASICs for each side are required, or four ASICs are needed to
 read out one \textit{FOXSI} DSSD.  
To connect from the DSSD to the ASICs, each DSSD strip is wire-bonded to the readout pads of the ASICs.  

\section{Basic Properties of a DSSD}
\begin{figure}[!t]
\centering
\includegraphics[width=3.5in]{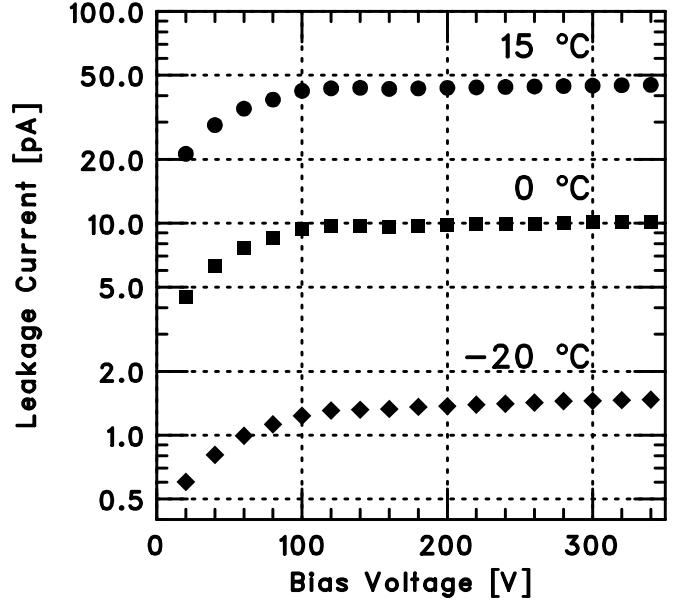}
\caption{I-V characteristic of \textit{FOXSI} DSSD.}
\label{fig_iv}
\end{figure}

We measured the leakage current of one central p-side strip of the \textit{FOXSI} DSSD 
at temperatures of 15$^\circ$C, 0$^\circ$C and $-$20$^\circ$C (Nominal in-flight temperature is $-$20$^\circ$C).  
Fig.~\ref{fig_iv} shows the measured $I$-$V$ characteristic.  
The leakage current is found to be proportional to $\exp (-1/kT)$.  
This is consistent with the idea that the origin of the current is thermal effect.
Under $-$20$^\circ$C, the leakage 
current is measured to be $\sim$1.5~pA with bias voltages of $>$200~V.
The corresponding ENC is $\sim$7~$e^-$ (RMS) at a shaping time of 3~$\mu$s calculated by a formula $\sim110\sqrt{I\tau}$~$e^-$ where $I$ is the leakage current in nA and $\tau$ is the time constant of a readout $CR$-$RC$ circuit in $\mu$s.
This result implies that the temperature of $-$20$^\circ$C is sufficiently low to obtain the best performance because the noise from the leakage current is negligible compared to normal readout noise.

\begin{figure}[!t]
\centering
\includegraphics[width=3.5in]{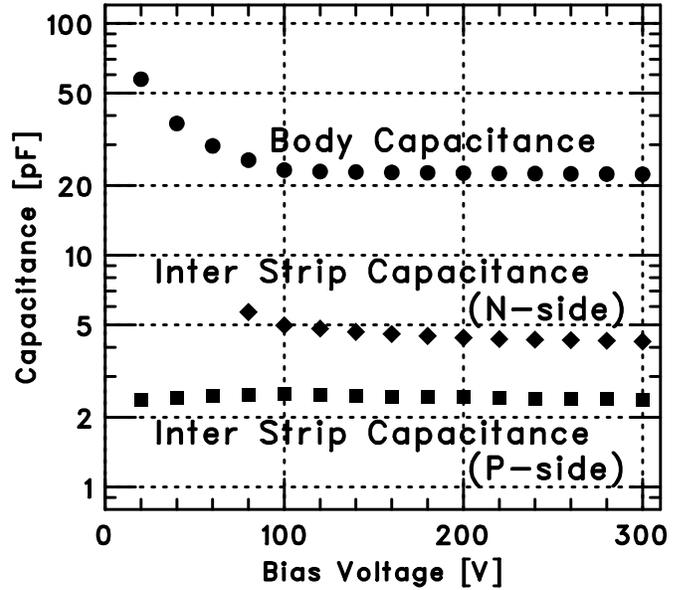}
\caption{C-V characteristic of \textit{FOXSI} DSSD.}
\label{fig_cv}
\end{figure}

Fig.~\ref{fig_cv} shows an C-V characteristic of the \textit{FOXSI} DSSD at a room temperature.  
The body capacitance and the inter-strip capacitances of p-side and n-side were measured.  
An inter-strip capacitance is a capacitance between one strip and the surrounding strips.  
The data of inter-strip capacitance are plotted for central p-side and n-side strips.
The amplifier noise due to the detector capacitance for a single strip can be parametrized as $\sqrt{20^2 + (10.5+7\times C_{\mbox{d}} )^2 + (12.2+8.1\times C_{\mbox{d}})^2/\tau}$~$e^-$, where $C_{\mbox{d}}$ is the sum of the body capacitance per single strip and the inter-strip capacitance.

The body capacitance becomes almost constant with bias voltages above $\sim$100~V, 
which implies that the DSSD is fully depleted at about 100~V.  
The inter-strip capacitance of the n-side can be measured above 80~V, and 
becomes almost constant above $\sim$200~V.  
This implies that n-side strips will be isolated above 80~V, and completely isolated above 
$\sim$200~V.  
The C-V characteristic suggests that a higher bias 
voltage such as 300~V is desired to reduce capacitance noise.  
The body capacitance for all strips is measured to be 22.4~pF, that is 0.2~pF per single strip with a bias voltage of 300~V.  
The inter-strip capacitances of the p-side and n-side are measured to be 2.3 and 4.2~pF at a bias voltage of 300~V.

\section{Performance of a Prototype for the \textit{FOXSI} Focal Plane Detector}
\begin{figure}[!t]
\centering
\includegraphics[width=3.5in]{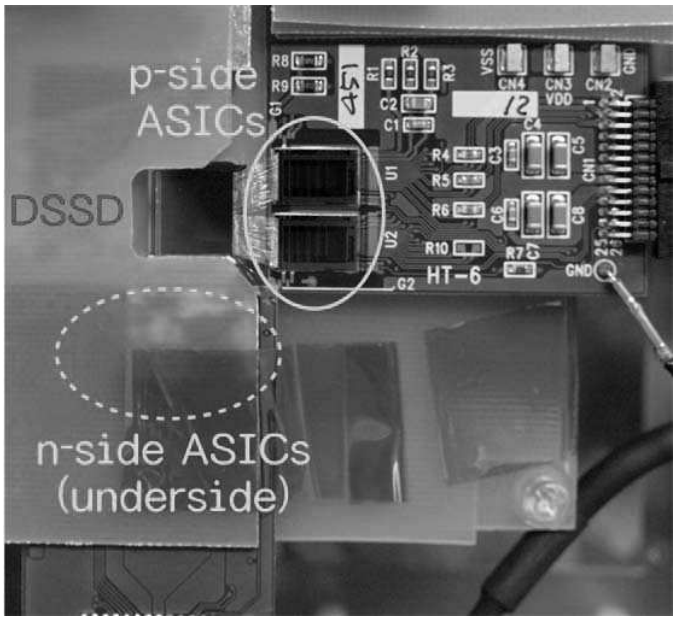}
\caption{Photo of the DSSD experimental setup.  Each strip of 
the DSSD is connected to an ASIC by wire-bonding.}
\label{fig_setup}
\end{figure}

We have fabricated a prototype of the \textit{FOXSI} focal plane detector that consists of one \textit{FOXSI} DSSD and four VA451 ASICs.
Fig.~\ref{fig_setup} shows a photo of the experimental setup.  
Wire-bonding from the detector to the ASICs was performed by a semi-automatic wire-bonding machine in our laboratory.  
Due to operational errors, some wires failed to be bonded.  
Therefore, some strips are not connected to the ASICs and cannot be read out.  
Such strips are electrically floated.  

Under a temperature of $-$20$^\circ$C and a bias voltage of 300~V, 
we successfully operated the \textit{FOXSI} DSSD.  
Events from 124 out of 128 strips of the p-side and 120 out of 128 strips for the n-side were used for the following analysis.  
In the spectral analysis, we use only events where a single hit is detected above the threshold of 5~keV in 
both the p-side and n-side in order to suppress
events with more than one interaction or any charge sharing between two adjacent strips.  

\begin{figure}[!t]
\centering
\includegraphics[width=3.5in]{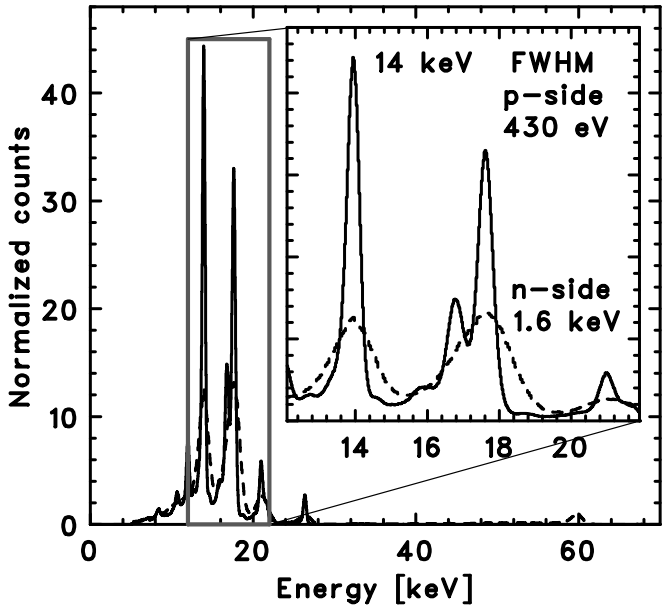}
\caption{Spectra from 
the p-side (solid line)
and n-side (dashed line) of 
the \textit{FOXSI} DSSD with 
a $^{241}$Am source.
The operating temperature was $-$20$^\circ$C and the bias voltage was 300~V.}
\label{fig_spec}
\end{figure}

Spectra from the p-side and n-side were obtained with the use of a $^{241}$Am radioactive isotope source.   
The spectra of all active strips on the p-side (black curve) and n-side (gray curve) are shown in Fig.~\ref{fig_spec}.
The energy resolutions of the p-side and n-side were measured to be 430~eV and 1.6~keV (FWHM) at 14~keV, respectively.
Since we can use the better energy information from the p-side for the spectral analysis, it is confirmed that the energy 
resolution fulfills the mission requirement.
\begin{figure}[!t]
\centering
\includegraphics[width=3.5in]{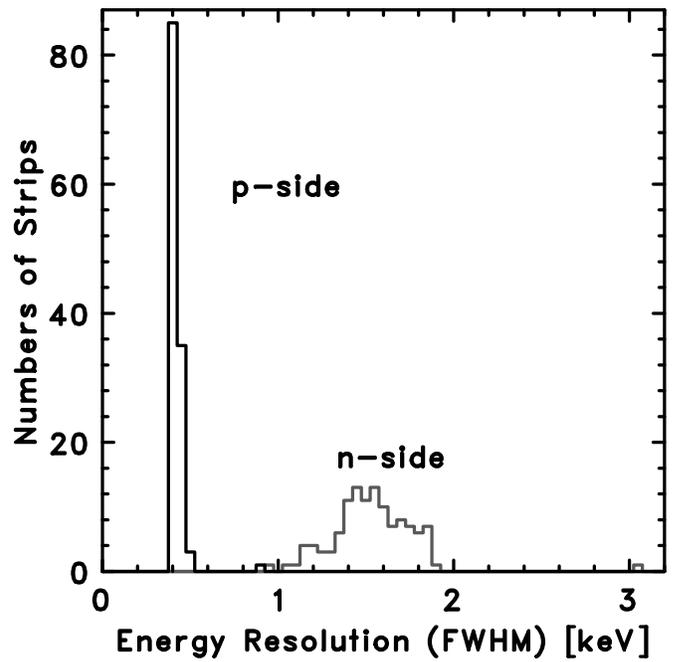}
\caption{Distribution of energy resolution (FWHM) for the 14~keV line.}
\label{fig_dedist}
\end{figure}

\begin{figure}[!t]
\centering
\includegraphics[width=3.5in]{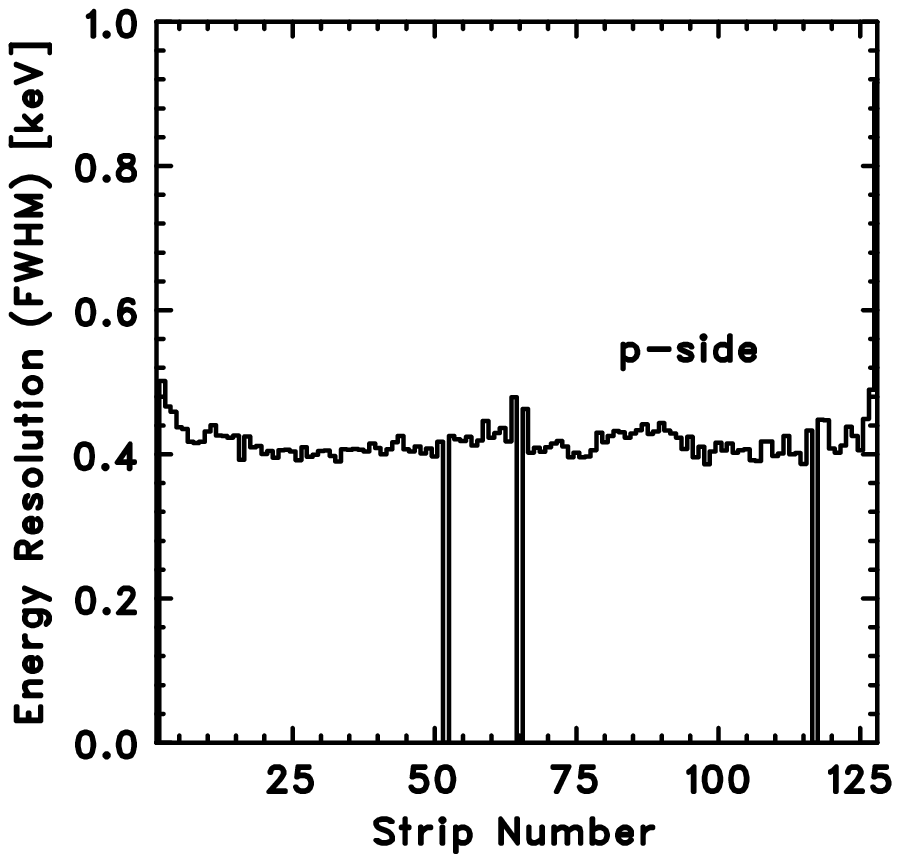}\\
\includegraphics[width=3.5in]{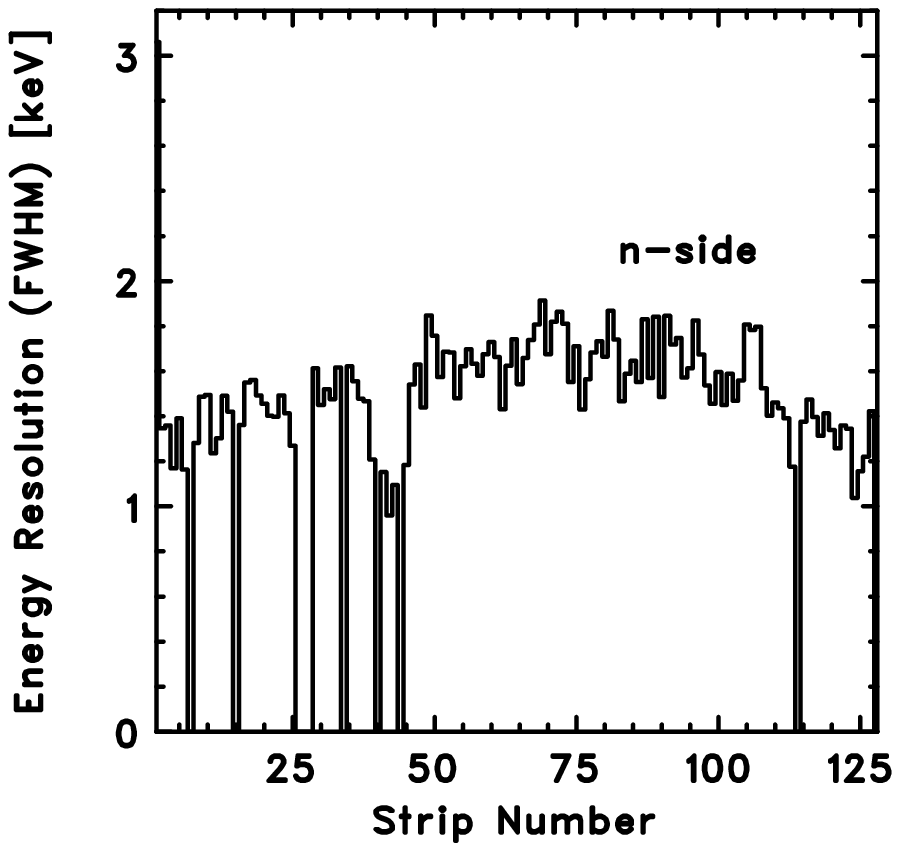}
\caption{Energy resolution (FWHM) for the 14~keV line as a function of the strip number.  Top: p-side.  Bottom: n-side.}
\label{fig:pos-dep}
\end{figure}

The distribution of energy resolutions (FWHM) of the p-side and n-side strips are shown in Fig.~\ref{fig_dedist}.
The energy resolutions of most channels are distributed from 0.4 to 0.5~keV on the p-side, excluding an exception with an energy resolution of 0.9~keV. 
The energy resolutions for the n-side are distributed between 0.9 and 1.9~keV with the exception of an outlier at 3.1~keV.  
Both outliers of the p-side and n-side correspond to edge strips, which neighbor the guard ring.  
Since the energy resolution and its dispersion are much worse on 
the n-side than on the p-side, the strip structure on the n-side is considered to be responsible for the excess noise on the n-side.
In fact, addition of the DC-coupled aluminum electrode on the p-stop improved the energy resolution.
A remaining noise source could be an accumulation layer between p-stops which may act as a floating n-strip with high resistance.
Fig.~\ref{fig:pos-dep} shows the position dependence of the energy resolution where we observe a moderate bump structure peaking 
on the n-side around strip numbers 70--90.
Since the resistance of the accumulation layer is highly dependent on the fabrication process, it can reasonably explain the asymmetric bump observed here.
On the other hand, we do not observe such structure or large dispersion in the energy resolution on n-side with another DSSD sample for the HXI fabricated in the same batch.
These observations may indicate that the excess noise on the n-side originates from the accumulation layer and may depend on the location in the silicon wafer and fabrication process.

\begin{figure}[t]
\centering
\includegraphics[width=3.4in]{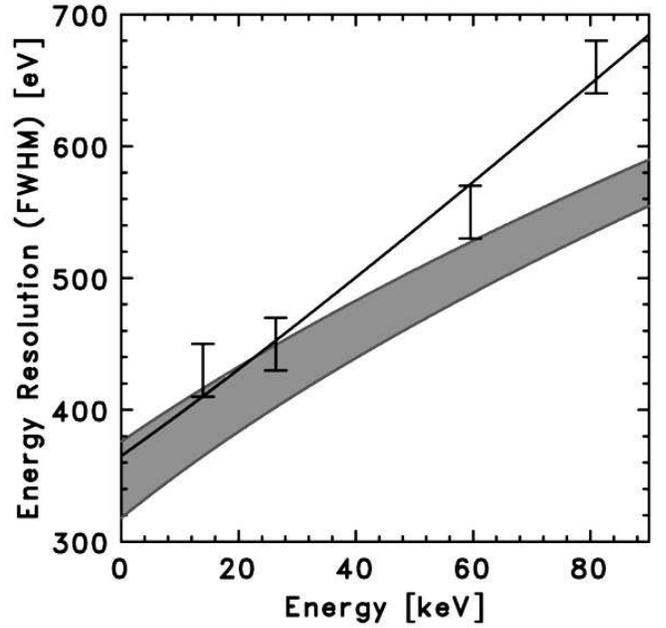}
\caption{Energy resolution (FWHM) as a function of the incident energy. Gray region indicate the energy resolution expected from the ASIC performance. Solid curve shows the fit result.}
\label{fig:energy-dep}
\end{figure}

Although the energy resolution observed for the 14~keV line on the p-side is much better than that on the n-side, it is still worse than the energy resolution expected from the noise performance of the ASIC, 
especially for higher energies.  
In order to study the nature of the excess noise on the p-side, we study the energy dependence of the energy resolution as shown in Fig.~\ref{fig:energy-dep}.
Data points are measured energy resolutions while a gray region gives the resolution calculated from 
the ASIC noise performance for a load with capacitance of ($3.0\pm0.5$)~pF at a peaking time of 4.8 ƒÊs ($346\pm28$~eV) 
and Fano noise (fluctuation of electron hole pairs produced by ionization) where the load capacitasnce includes a parasitic capacitancer of $(0.5\pm0.5)$~pF.  
A fit to an empirical formula, $\sqrt{E_0^2+(f_1E)^2+N_\mathrm{Fano}(E)^2}$ yields $E_0=365\pm15$~eV, $f_1=(4.0\pm0.5)\times10^{-3}$ where $N_\mathrm{Fano}(E)$ is the Fano noise.
The linear term may be explained by the gain uncertainties of the ASIC as mentioned in previous papers\cite{tajima2003, tajima2004vata}.

While we can achieve a spectral resolution better than $<$0.5~keV in the \textit{FOXSI} energy range solely due to a good noise performance on 
the p-side, the localization of the incident X-ray position still relies on detecting X-ray signals on both the p-side and the n-side.  
Since the noise performance on the n-side is marginal, we calculate the probability for wrong localizations.
Assuming a Gaussian noise distribution, the energy resolution of 2.0~keV for the n-side corresponds to a chance probability of $<$10$^{-7}$ 
for recording an energy above 5~keV due to noise alone.
This implies that positions of incident photons can be determined with a negligible probability of spurious hits.
Therefore, we confirm that the DSSD can be operated as an imager at a low energy threshold of 5~keV, which fulfills the scientific requirements of the \textit{FOXSI} mission.

\begin{figure}[!t]
\centering
\includegraphics[width=3.5in]{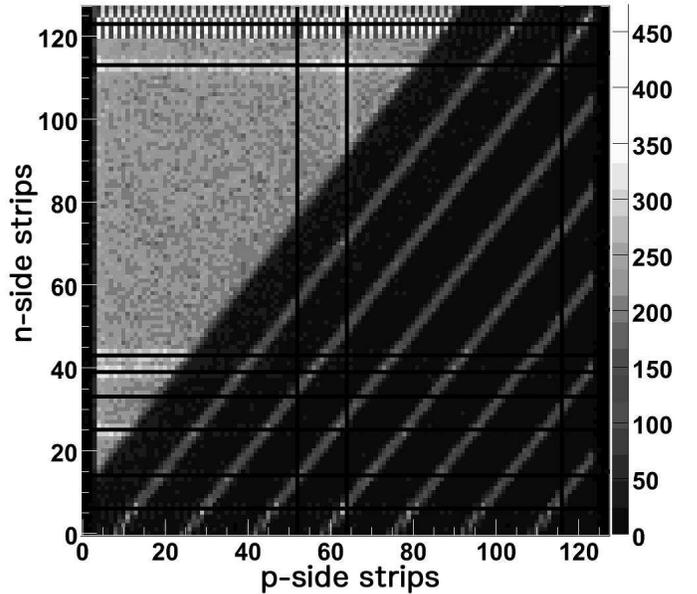}
\caption{Shadow image from the
\textit{FOXSI} DSSD with a radio isotope of $^{133}$Ba.  The scale gives the number of  counts.  
   The width of the slits in this image is 100~$\mu$m.}
\label{fig_image}
\end{figure}

Fig.~\ref{fig_image} shows a shadow image of a tungsten plate with 1~mm pitch slits using X-rays from a $^{133}$Ba source in an energy band from 20 to 40~keV.
To test the imaging performance, the tungsten plate is tilted by 45$^{\circ}$ and does not cover the top left part of the detector.  
The width of the slits is 100~$\mu$m.  
The individual slits are clearly visible, successfully demonstrating the capabilities of fine-pitch imaging and spectroscopy.

In the left part of the top edge of the image ($x\le$~90 and $y\ge$~120), a longitudinal stripe pattern appears even though counts should be almost flat in this region since there is no mask.
In the edges of the p-side strips (the top edge and the bottom edge of the image), 
wire-bonding pads are placed, 
 which is considered to be the cause of the pattern.
The bonding pads are wider than the strips, and therefore staggered. 
Events in this region can be used for spectroscopy although the spatial resolution for the $x$-position is worse than the other 
regions.  

On the other hand, although signals are successfully obtained by the p-side, 
the right and left edges on the image ($x\le$~2 and $x\ge$~125) have no n-side signal.  
This is considered to be caused by the presence of wire-bonding pads on n-side.  
In this region, events can be used only for spectroscopy.  

\section{Conclusions}
\textit{FOXSI} is a sounding rocket mission which will
observe the Sun in the hard X-rays (5-15~keV) 
by using HXR focusing optics and a semiconductor imaging detector. 
We have developed fine-pitch and low noise DSSDs for the \textit{FOXSI} focal plane detector, specifically to fulfill 
scientific requirements on the spatial resolution, energy resolution, low threshold energy and time resolution.
We designed and fabricated a DSSD with a thickness of 500~$\mu$m, and a dimension of 9.6$\times$9.6~mm 
containing 128 strips 
on a pitch of 75~$\mu$m, which corresponds to 8~arcsec at the focal length of 2~m.  
The DSSD was successfully operated in a laboratory experiment under a temperature of $-$20$^\circ$C and a bias voltage of 300~V.  
The energy resolution was measured to be 430~eV for the p-side and 1.6~keV for the n-side at 14~keV, sufficient for the \textit{FOXSI} mission requirement.  
We also successfully obtained a shadow image and thus demonstrated the capabilities of both fine-pitch imaging and spectroscopy.  

\bibliography{ishikawa2010}
\bibliographystyle{IEEEtran}

% that's all folks
\end{document}